\documentclass[11pt,a4paper]{article}

\usepackage[UKenglish]{babel}
\usepackage[T1]{fontenc}
\usepackage{lmodern}
\usepackage{microtype}

\usepackage{amsmath,amsthm,amssymb,mathtools}
\usepackage{mathrsfs}
\usepackage{tikz}
\usepackage{color}

\usepackage{hyperref}
\usepackage{url}
\usepackage{comment}
\usepackage{thmtools}
\usepackage{thm-restate}
\usepackage{caption}
\usepackage[noend]{algpseudocode}
\usepackage[ruled,vlined,boxed]{algorithm2e}
\usepackage{enumitem}
\usepackage{geometry}
\theoremstyle{plain}
\newtheorem{theorem}{Theorem}[section]
\newtheorem{lemma}[theorem]{Lemma}

\newtheorem{claim}[theorem]{Claim}
\newtheorem{corollary}[theorem]{Corollary}
\newtheorem{fact}[theorem]{Fact}
\newtheorem{defi}[theorem]{Definition}
\theoremstyle{definition}
\newtheorem{definition}[theorem]{Definition}

\theoremstyle{remark}
\newtheorem{remark}[theorem]{Remark}

\newcommand{\pmone}{\{-1,+1\}}
\newcommand{\ftwo}{\mathbb{F}_2}

\newcommand{\cL}{\mathcal{L}}

\newcommand{\Var}{\mathrm{Var}}

\renewcommand{\hat}{\widehat}
\renewcommand{\tilde}{\widetilde}

\newcommand{\dA}{\delta_{\cL}}

\newcommand{\sign}{\mathsf{sign}}

\title{\bfseries
Spectral Norm, Economical Sieve, and Linear Invariance Testing of Boolean Functions
}

\author{
Swarnalipa Datta
\footnote{Indian Statistical Institute, Kolkata, India}
\and
Arijit Ghosh
\footnotemark[1]
\and
Chandrima Kayal
\footnote{Universit\'e Paris Cit\'e, CNRS, IRIF, Paris, France}
\and
Manaswi Paraashar
\footnote{University of Copenhagen, Copenhagen, Denmark}
\and 
Manmatha Roy
\footnotemark[1]
}

\date{}

\begin{document}
\maketitle

\begin{abstract}
Given Boolean functions \( f, g : \mathbb{F}_2^n \to \{-1,+1\} \), we say they are {\em linearly isomorphic} if there exists \( A \in \mathrm{GL}_n(\mathbb{F}_2) \) such that \( f(x)=g(Ax) \) for all \( x \). We study this problem in the tolerant property testing framework under the known--unknown model, where \( g \) is given explicitly and \( f \) is accessible only via oracle queries, meaning the algorithm may adaptively request the value of \( f(x) \) for inputs \( x \in \mathbb{F}_2^n \) of its choice. Given parameters \( \epsilon \ge 0 \) and \( \omega>0 \), the goal is to distinguish whether there exists \( A \in \mathrm{GL}_n(\mathbb{F}_{2})\) such that the normalized Hamming distance between \( f \) and \( g(Ax) \) is at most \( \epsilon \), or whether for every \( A \in \mathrm{GL}_n(\mathbb{F}_2) \) the distance is at least \( \epsilon+\omega \).

Our main result is a tolerant tester making \( \widetilde{O} \left( \left( m/\omega \right)^4 \right) \) queries to \( f \), where \( m \) is an upper bound on the spectral norm of \( g \), improving the previous \( \widetilde{O} \left( \left( m/\omega \right)^{24} \right) \) bound of Wimmer and Yoshida. We complement this with a nearly matching lower bound of \( \Omega(m^2) \) for constant \( \omega \) (for example, \( \omega=1/4 \)), improving the prior \( \Omega(\log m) \) lower bound of Grigorescu, Wimmer and Xie. A key technical ingredient on the algorithmic side is a query-efficient local list corrector. For the lower bound, we give a reduction from communication complexity using a novel subclass of Maiorana--McFarland functions from symmetric-key cryptography.

\end{abstract}

\newpage

\section{Introduction}
\label{sec: intro}


A Boolean function \( f : \mathbb{F}_2^n \to \{-1,+1\} \) is said to be linearly isomorphic to
\( g : \mathbb{F}_2^n \to \{-1,+1\} \) if there exists \( A \in \mathrm{GL}_n(\mathbb{F}_2) \) such that
\( f(x) = g(Ax) \) for all \( x \in \mathbb{F}_2^n \).
The linear isomorphism is a fundamental notion with applications in coding theory, circuit complexity,
and cryptography~\cite{CV03,BA04,YB12,ND06,DT01,XH94,CC07,OD12}.
A notable example is given by Reed--Muller codes, which are affine-invariant and whose decoding algorithms
crucially exploit this symmetry~\cite{Abbe2020-aw}.
Closely related isomorphism problems also arise in cryptography, such as the
\emph{Isomorphism of Polynomials} problem~\cite{poly06}.


From an algorithmic standpoint, this problem has been extensively studied. Determining whether two Boolean functions are linearly isomorphic is computationally intractable even under the strong restriction that the linear transformation is a permutation matrix. The problem is known to be coNP-hard (when the functions are given in DNF form) and lies in \( \Sigma_2^p \), but it is not known to be in coNP. Agrawal and Thierauf further showed that, unless the polynomial hierarchy collapses to \( \Sigma_3^p \), the problem is not \( \Sigma_2^p \)-complete~\cite{agariso}. The best known algorithm reduces the task to the \emph{Hypergraph Isomorphism Problem} and runs in \( 2^{O(n)} \) time~\cite{acmHypergraphIsomorphism}.


In this work, we study linear isomorphism of Boolean functions in the \emph{property testing} framework,
where the goal is to decide whether an object has a given property or is far from having it, using only a small number of queries.
We work in the stronger \emph{tolerant testing} model, which requires distinguishing objects that are close to the property from those that are far.
We now formally define the problem of testing linear isomorphism in this setting.





\begin{definition}[\textsc{$(\epsilon,\omega)$-Tolerant Linear Isomorphism Testing}]
Let \( \epsilon \ge 0 \) and \( \omega > 0 \).
An algorithm is given full access to a known function
\( g : \mathbb{F}_2^n \to \{-1,+1\} \) and oracle access to an unknown function
\( f : \mathbb{F}_2^n \to \{-1,+1\} \), meaning that for any input
\( x \in \mathbb{F}_2^n \) the algorithm may query the oracle to obtain \( f(x) \).
It must distinguish, with probability at least \(2/3\), between the following cases:
\begin{itemize}
    \item
        {\rm {\bf (Yes)}} There exists \( A \in \mathrm{GL}_n(\mathbb{F}_2) \) such that
    \( \delta(f \circ A, g) \le \epsilon \).
    \item
        {\rm {\bf (No)}} For all \( A \in \mathrm{GL}_n(\mathbb{F}_2) \),
    \( \delta(f \circ A, g) \ge \epsilon + \omega \).
\end{itemize}
Here \( \delta(f,g) \) denotes the fraction of inputs on which \( f \) and \( g \) differ, that is,
\( \delta(f,g) := \mathop{\Pr}_{x \sim \mathbb{F}_2^{n}} \left[ f(x) \neq g(x) \right] \).
\end{definition}
We now review prior work on the isomorphism testing problem, covering both algorithmic upper bounds and lower bounds.

\subsection{Related Works}

Wimmer and Yoshida~\cite{WY13} were the first to study linear isomorphism testing in the tolerant setting.
They gave an algorithm for the \((\omega,3\omega)\)-tolerant problem with query complexity
\( \widetilde{O}((m/\omega)^{24}) \), where \( m \) is an upper bound on the spectral norm of the known function \( g \).
They also proved an adaptive lower bound of \( \Omega(m) \) queries in a certain subconstant regime of \( \omega \) (see Remark~\ref{remark:Wimmer_and_Yoshida_lower_bound}).
In addition, Chakraborty et al.~\cite{ChakrabortyFGM12} showed an \( \Omega(k) \) lower bound for testing linear isomorphism, when the known function is a \( k \)-Junta,
which implies an \( \Omega(\log m) \) lower bound in terms of the spectral norm of the known function.
Grigorescu, Wimmer, and Xie~\cite{GW13} further proved an adaptive lower bound of \( \Omega(n^2) \) queries
when the known function \( g \) is the inner-product function on \( n \) bits.
Since the spectral norm of the inner-product function is at most \( 2^{n/2} \), this also yields an
\( \Omega(\log m) \) lower bound in terms of \( m \).

Linear isomorphism is an affine-invariant property, and testing affine-invariant properties of Boolean functions
has been widely studied; see, e.g.,~\cite{KS08,BGS15,BFHH13,regularityfunction}.
In particular, the {\em regularity framework} of Hatami and Lovett~\cite{regularityfunction} yields tolerant testers
for a broad class of affine-invariant properties, including linear isomorphism to a fixed function.
However, a major limitation of these general frameworks is that the resulting query complexity grows as a
tower-type function of the relevant complexity parameter.

Characterizing which properties admit constant-query testers is a central question in property testing.
In graph property testing, a landmark result is the characterization of all testable properties via
Szemer\'{e}di’s regularity lemma~\cite{alon2006combinatorial}.
In our setting, Wimmer and Yoshida~\cite{WY13} gave an analogous characterization for linear isomorphism testing,
showing that functions with small spectral norm are exactly those for which linear isomorphism can be tested
with a constant number of queries.

A closely related and more restricted problem is testing isomorphism under permutations of variables,
which has also been extensively studied~\cite{BlaisO10,AlonB10,ChakrabortyGM11,ChakrabortyFGM12,BlaisWY15,BlaisCELR19}.
Alon et al.~\cite{AlonBCGM13} showed that testing permutation isomorphism for arbitrary Boolean functions
requires \( \Omega(n) \) queries and can be done with \( O(n \log n) \) queries.
They also proved that testing permutation isomorphism to a \( k \)-junta requires \( \Omega(k) \) queries,
and also gave a nearly matching \( O(k \log k) \)-query algorithm.

\subsection{Our Contribution}

Prior work exhibits a large gap between the known upper and lower bounds, even for Boolean functions
with small spectral norm, which form the only class admitting constant-query testers.
In this work, we nearly close this gap by presenting a non-adaptive randomized tester with
substantially improved query complexity for all \( \epsilon \ge 0 \) and \( \omega > 0 \).

\begin{restatable}{theorem}{upperboundtheorem}
\label{theorem: lineariso testing upperbound}
Let \( \epsilon \ge 0 \) and \( \omega > 0 \).
There exists a non-adaptive randomized algorithm (see Algorithm~\ref{algo: tester for lineariso})
that solves the \textsc{$(\epsilon, \omega)$-Tolerant Linear Isomorphism Testing} problem for a known function
\( g : \mathbb{F}_2^n \to \{-1, +1\} \) with spectral norm at most \( m \) and an unknown function
\( f : \mathbb{F}_2^n \to \{-1, +1\} \), using \( \widetilde{O}((m/\omega)^4) \) queries to \( f \),
and succeeding with probability at least \( 2/3 \).
Here, the \( \widetilde{O}(\cdot) \) notation hides polynomial factors in \( \log m \) and \( \log(1/\omega) \).
\end{restatable}

We complement this upper bound with the following nearly matching lower bound.

\begin{restatable}{theorem}{lowerboundtheorem}
\label{theorem: lineariso testing lowerbound}
For any \( m > 0 \), there exists a Boolean function
\( h : \mathbb{F}_2^n \to \{-1, +1\} \) with spectral norm at most \( m \)
such that every adaptive algorithm for the
\textsc{$(0, 1/4)$-Tolerant Linear Isomorphism Testing} problem with respect to \( h \)
requires \( \Omega(m^2) \) queries to the unknown function.
\end{restatable}

\begin{remark}
\label{remark:Wimmer_and_Yoshida_lower_bound}
Wimmer and Yoshida~\cite{WY13} proved an \( \Omega(m) \) lower bound for the
\( (0, \omega) \)-tolerant linear isomorphism testing problem.
However, to the best of our understanding, their proof applies only to the special case \( \omega = 1/m \).
Even if their argument could be extended to general \( \omega \), our lower bound is quadratically stronger
and is based on a different approach, which we believe may be of independent interest.
\end{remark}

\section{Proof ideas}

\subsection{Proof idea of Theorem~\ref{theorem: lineariso testing upperbound}}

At a high level, we follow the paradigm of testing by implicit learning introduced by
Diakonikolas et al.~\cite{diakonikolas2007testing}, and build on the approach of
Gopalan et al.~\cite{gopalan2011testing} for testing induced membership in subclasses of Boolean functions with small spectral norm,
which was also used by Wimmer and Yoshida~\cite{WY13} for linear isomorphism testing.
Our main contribution is a refinement of this approach that yields a tester with significantly improved
query complexity, comapred to Wimmer and Yoshida~\cite{WY13}. In the process, we introduce an efficient list-correction framework, which we call the
\emph{Economical Sieve}, and which we believe may be of independent interest in the analysis of Boolean functions.
We now outline the overall proof strategy.

We first recall that if two Boolean functions are approximately related by a linear transformation,
then their Fourier spectra are strongly correlated.
Formally, for \( f, g : \mathbb{F}_2^n \to \{-1,1\} \), if there exists an invertible matrix
\( A \in \mathrm{GL}_n(\mathbb{F}_2) \) such that \( \delta(f, g \circ A) \le \epsilon \), then
\[
\sum_{\beta \in \mathbb{F}_2^n} \widehat f(\beta)\, \widehat{g \circ A}(\beta) \;\ge\; 1 - 2\epsilon,
\]
and the converse also holds.
Moreover, when the known function \( g \) has small spectral norm, this implies that it suffices to focus
on the heavy Fourier coefficients of the unknown function \( f \).
The main challenge is to locate and estimate these coefficients using only oracle access to \( f \),
with a number of queries independent of \( n \).
At first glance, this seems difficult, since even identifying a linear function requires
\( \Omega(n) \) queries; see, e.g.,~\cite[Exercise~1.27]{ODonnellbook2014}.

We construct an approximation \( f^* \) of the unknown function \( f \) such that there exists (yet unknown to the algorithm)
\( A \in \mathrm{GL}_n(\mathbb{F}_2) \) such that
$
\| f - f^* \circ A \|_2^2 \le \epsilon,
$
using \( \mathrm{poly}(m,1/\epsilon) \) queries to \( f \), where \( m \) is the spectral norm of the known
function \( g \).
The crucial observation is that, for testing linear isomorphism, it suffices to learn \( f \) only up to
an unknown nonsingular linear transformation.

The key tool enabling this is a local list-correction algorithm.
Intuitively, given a threshold parameter \( \theta \), oracle access to \( f \), and a point
\( x \in \mathbb{F}_2^n \), the algorithm identifies all heavy Fourier coefficients of \( f \) and returns a list
$
L_x = \left\{ \chi_\beta(x) : |\widehat f(\beta)| \ge \theta \right\}.
$
The algorithm is not required to output the Fourier characters \( \beta \) themselves, but only their
evaluations at the given point \( x \).
Its query complexity depends only on \( \theta \) and is independent of the dimension \( n \).
For our purposes, this weaker guarantee is sufficient. By accessing \( L_x \) at many random inputs \( x \) (polynomially many in \( 1/\theta \)), we show how one can (i) approximate the heavy Fourier coefficients of \( f \), and (ii) recover a basis for them, thereby identifying the coefficients up to a linear transformation. This yields a function that approximates \( f \) up to linear isomorphism, which is adequate for our setting, since linear isomorphism is itself a linear-invariant property. Thus, it suffices to check over all possible linear transformations to determine whether a suitable one exists. This can be carried out offline and does not require any further queries to the unknown function \( f \). It remains to describe the local list-correction algorithm, which we refer to as the
\textsf{Economical Sieve}.

The \textsf{Economical Sieve} can be viewed as a query-efficient refinement of the implicit learning framework
known as the \emph{Implicit Sieve}, introduced by Wimmer and Yoshida~\cite{WY13}.
Our improved query complexity is achieved through a more refined analysis of the
\emph{coset hashing} process.
Specifically, we revisit coset hashing process and establish new concentration bounds
(Claim~\ref{lemma_bucket_discard} and~\ref{lem:projection_concentration})
for the \( \ell_1 \)- and \( \ell_2 \)-norms of the projected Fourier spectrum.
These results show that, within any coset, the norm of the Fourier projection is dominated by the
contribution of heavy coefficients. Our analysis is inspired by techniques for heavy-hitter detection in the streaming literature:
\( \ell_2 \)-concentration plays a role analogous to the \textsc{Count--Min Sketch}, while
\( \ell_1 \)-concentration corresponds to the \textsc{Count Sketch}.
Below we provide a brief sketch of the overall approach.

\begin{itemize}

\item \textbf{Heavy Fourier Coefficient Detection.}  
We begin by projecting the Fourier spectrum of \( f \) onto cosets of a random subspace
\( H \subseteq \mathbb{F}_2^n \) of dimension \( \log O(1/\theta^8) \).
This induces a pairwise-independent hashing of the Fourier coefficients of \( f \) into
\( O(1/\theta^{8}) \) cosets of \( H \). The key ingredient is Claim~\ref{lemma_bucket_discard}, which shows that the
\( \ell_2 \)-norm of the Fourier projection in each coset is dominated by its heaviest coefficient.
Thus, given sufficiently accurate \( \ell_2 \)-estimates for the cosets, we can identify all cosets
containing a heavy Fourier coefficient. Unlike prior work~\cite{WY13,gopalan2011testing}, which relied on estimating the Fourier
$\ell_4$-norm of coset buckets in order to detect cosets containing heavy Fourier
coefficients, our method relies solely on these new concentration results. Consequently,
we avoid the $\ell_4$-estimation step, which is inherently query-intensive.



\item \textbf{Heavy Fourier Coefficient Evaluation.}  
Once the heavy cosets have been identified, we proceed to evaluate the corresponding heavy
Fourier characters. This step relies on Claim~\ref{lem:projection_concentration}, which shows that the \( \ell_1 \)-norm
of the projected Fourier spectrum within each coset is dominated by its largest coefficient.
Consequently, for any \( x \in \mathbb{F}_2^n \), the projection can be well approximated by the
evaluation of the heaviest Fourier coefficient at \( x \). However, since we use much smaller coset structures than in~\cite{WY13}, the probability that this
approximation is sufficiently accurate for a fixed \( x \) is only constant. As our algorithm requires many such evaluations, we apply an amplification step in the spirit of the Goldreich--Levin algorithm~\cite{goldreich1989hard}.

\end{itemize}

\subsection{Proof idea of Theorem~\ref{theorem: lineariso testing lowerbound}}

We prove our lower bound for linear isomorphism testing via a reduction from the
\emph{Approximate Matrix Rank} problem in the public-coin randomized communication complexity model.
Our reduction follows the general communication-based framework for proving lower bounds in function
property testing introduced by Blais, Brody and Matulef~\cite{ptlbcomm}.
A key ingredient in our approach is the structural relationship between the sparsity of a Boolean
function’s Fourier support and the dimension of its linear span, known as the \emph{Fourier dimension}.
A fundamental result of Sanyal~\cite{sanyal2019fourier} shows that the Fourier dimension is at most the
square root of the Fourier sparsity, up to constant factors.

To exploit this relationship, we construct a class of Boolean functions based on the
Maiorana--McFarland construction, which is widely used in symmetric-key cryptography due to its rich
algebraic structure and well-understood spectral properties.
By composing these functions with linear maps of varying rank, we obtain a family of functions whose
Fourier sparsity and Fourier dimension can be carefully controlled, while preserving the quadratic
relationship established in~\cite{sanyal2019fourier}.
Depending on the rank of the applied linear transformation, functions from this family are either
linearly isomorphic to one another or far from any such isomorphism.
This enables a reduction from the \emph{Approximate Matrix Rank} problem to testing linear isomorphism
of Boolean functions.

In this reduction, Alice and Bob are given matrices \( A \) and \( B \), respectively, and must decide
whether the rank of their sum \( C = A + B \) is exactly \( r \), or significantly smaller, using limited
communication and shared public randomness.
We encode these matrices into Boolean functions from our constructed family such that:
\begin{itemize}
    \item if \( A + B \) has full rank, the resulting joint function is a \emph{yes}-instance of the
    linear isomorphism testing problem;
    \item if \( A + B \) is far from full rank (by a constant factor), the resulting joint function is a
    \emph{no}-instance.
\end{itemize}

To complete the reduction, we use a result of Sherstov and Storozhenko~\cite{matrixrank}, which proves an
\( \Omega(r^2) \) lower bound on the randomized communication complexity of distinguishing whether a
matrix has rank \( r \) or at most \( c r \), for a constant \( c < 1 \).
This lower bound translates directly into a nearly matching query lower bound for linear isomorphism testing
in our construction.


\section{Background}
\label{section: background}

In this section, we recall standard notions from the analysis of Boolean functions and
communication complexity that will be used throughout the paper.

For \( \alpha, \beta \in \mathbb{F}_2^{n} \), \( \langle \alpha, \beta \rangle \) denotes the
\( \mathbb{F}_2 \)-inner product.
For a function \( f : \mathbb{F}_2^{n} \to \mathbb{R} \), we write
\[
\mathop{\mathbb{E}}_{x \sim \mathbb{F}_2^{n}}[ f(x) ]
:= \frac{1}{2^{n}} \sum_{x \in \mathbb{F}_2^{n}} f(x).
\]

For a function \( f : \mathbb{F}_2^n \to \mathbb{R} \), its Fourier coefficient at
\( \alpha \in \mathbb{F}_2^n \) is
\[
\widehat f(\alpha) := \mathop{\mathbb{E}}_{x \sim \mathbb{F}_2^n}[f(x)\chi_\alpha(x)],
\qquad
\text{where } \chi_\alpha(x) := (-1)^{\langle \alpha, x \rangle}.
\]
All expectations and probabilities over \( \mathbb{F}_2^n \) are with respect to the uniform
distribution.
Every such function admits the Fourier expansion
\[
f(x) = \sum_{\alpha \in \mathbb{F}_2^n} \widehat f(\alpha)\chi_\alpha(x).
\]

\begin{fact}[Plancherel and Parseval Identities, see~{\cite[Chapter~1]{ODonnellbook2014}}]
\label{fact: parseval}
For any functions \( f, g : \mathbb{F}_2^n \to \mathbb{R} \),
\[
\mathop{\mathbb{E}}_{x \sim \mathbb{F}_2^{n}}[f(x) g(x)]
= \sum_{\alpha \in \mathbb{F}_2^n} \widehat f(\alpha)\widehat g(\alpha).
\]
In particular,
\[
\mathop{\mathbb{E}}_{x \sim \mathbb{F}^{n}_{2}}[f(x)^2] = \sum_{\alpha \in \mathbb{F}_2^n} \widehat f(\alpha)^2,
\]
which equals \(1\) when \( f : \mathbb{F}_2^n \to \{-1,1\} \).
\end{fact}

The \emph{Fourier support} of \( f \) is
\( \mathrm{supp}(\widehat f) := \{\alpha \in \mathbb{F}_2^n : \widehat f(\alpha) \neq 0\} \),
and its size is called the \emph{Fourier sparsity}.
We say that \( f \) is \( s \)-Fourier sparse if \( |\mathrm{supp}(\widehat f)| \le s \).
We also use the \emph{spectral norm} of \( f \), defined as
\( \|\widehat f\|_1 := \sum_{\alpha} |\widehat f(\alpha)| \).

For Boolean functions \( f, g : \mathbb{F}_2^n \to \{-1,1\} \), we measure distance by
\[
\delta(f,g) := \Pr_{x \sim \mathbb{F}_2^n}[f(x) \neq g(x)].
\]
The \emph{Linear Isomorphism Distance} is
\[
\dA(f,g) := \min_{A \in \mathrm{GL}_n(\mathbb{F}_2)} \delta(f \circ A, g),
\]
where \( \mathrm{GL}_n(\mathbb{F}_2) \) denotes the group of invertible linear maps over
\( \mathbb{F}_2 \).
Both \( \delta \) and \( \dA \) satisfy the triangle inequality.

\begin{fact}[Self-Correction, see~{\cite[Proposition~1.31]{ODonnellbook2014}}]
\label{lemma:self-correction}
Suppose \( f : \mathbb{F}_2^n \to \{-1,1\} \) is \( \epsilon \)-close to the linear function
\( \chi_\alpha \).
Then, for every \( x \in \mathbb{F}_2^n \),
\[
\Pr_{y \sim \mathbb{F}_2^n}\bigl[\chi_\alpha(x) = f(y)f(x+y)\bigr] \ge 1 - 2\epsilon .
\]
\end{fact}

For the lower bound, we rely on a communication complexity result for the
\emph{Approximate Matrix Rank} problem.
In this problem, Alice receives a matrix \( A \in \mathbb{F}_2^{r \times r} \) and Bob receives
\( B \in \mathbb{F}_2^{r \times r} \); their goal is to distinguish whether
\( \operatorname{rank}(A+B) = r \) or \( \operatorname{rank}(A+B) \le r/4 \).
We use the following fundamental result of Sherstov and Storozhenko~\cite{matrixrank}.

\begin{fact}[Approximate Matrix Rank Problem~{\cite[Theorem~1.1]{matrixrank}}]
\label{theorem: actual rank}
Alice receives a matrix \( A \in \mathbb{F}_2^{r \times r} \), and Bob receives a matrix
\( B \in \mathbb{F}_2^{r \times r} \).
Their task is to determine, for \( C = A + B \), whether
\[
\operatorname{rank}(C) = r
\quad \text{or} \quad
\operatorname{rank}(C) \le \tfrac{r}{4},
\]
while exchanging as few bits of communication as possible.
The public-coin randomized communication complexity of this problem is
\( \Omega(r^2) \).
\end{fact}

\section{A query-efficient tester for the linear isomorphism problem }
\label{section: testing upper bound}

We begin with the local list correction framework, which plays a fundamental role in our algorithm.

\begin{lemma}[\textsf{Economical Sieve}]
    \label{lemma: economicsieve}
There exists an algorithm, \textsf{Economical Sieve}, that takes parameters \( \theta \) and \( \lambda \) as input, makes 
$ \widetilde{O} \left( \max\left( \frac{1}{\theta^4}, \frac{\lambda}{\theta^2} \right) \right)$ 
queries to the truth table of a Boolean function \( f: \mathbb{F}_2^n \to \{-1, 1\} \), and, with probability at least \( \frac{9}{10} \),  outputs: 

\begin{itemize}
    \item A matrix \( Q \in {\pmone}^{\lambda \times k} \), where the \( (i,j) \)-th entry is \( \chi_{\alpha_j}(x_i) \);
    \item A column vector \( F \in \{-1,1\}^\lambda \), where \( F(i) = f(x_i) \) for each \( i \in [\lambda] \),
\end{itemize}
where each \( x_i \in \mathbb{F}_2^n \) is drawn independently and uniformly at random, and each \( \alpha_j \in \mathbb{F}_2^n \) belongs to a set \( \mathcal{S} = \{ \alpha_1, \dots, \alpha_k \} \) satisfying:

\begin{itemize}

    \item For every \( \alpha \in \mathbb{F}_2^n \) such that \( |\widehat{f}(\alpha)| \geq \theta \), we have \( \alpha \in \mathcal{S} \);
    \item For all \( \alpha \in \mathcal{S} \), it holds that \( |\widehat{f}(\alpha)| \geq {\theta}/{2} \).
\end{itemize}
\end{lemma}

For now, we assume Lemma~\ref{lemma: economicsieve} and proceed to prove Theorem~\ref{theorem: lineariso testing upperbound}. The proof of Lemma~\ref{lemma: economicsieve}, which establishes the correctness and performance guarantees of the \textsf{Economical Sieve}, will be presented later.

\subsection{Proof of Theorem \ref{theorem: lineariso testing upperbound}}
We begin by restating Theorem \ref{theorem: lineariso testing upperbound} for the sake of reader's convenience. 

\upperboundtheorem*

\begin{proof}

We prove the theorem by presenting an explicit algorithm, \textsf{LinearIsoTester} (Algorithm~\ref{algo: tester for lineariso}), and analyzing its theoretical guarantees to show that it satisfies the requirements of Theorem~\ref{theorem: lineariso testing upperbound}. Consider Algorithm~\ref{algo: tester for lineariso}. We assume that, for the parameter settings 
\[
\theta = \frac{\omega}{10 m} \quad \text{and} \quad \lambda = \frac{100}{\theta^2} \log \frac{100}{\theta^2},
\] 
Lemma~\ref{lemma: economicsieve} holds and the \textsf{Economical Sieve} correctly returns the pair $(Q, F)$. We now analyze the subsequent steps of Algorithm~\ref{algo: tester for lineariso}. Recall that each invocation of the \textsf{Economical Sieve} implicitly returns a set 
\(
\mathcal{S} = \{ \alpha_1, \dots, \alpha_k \} \subseteq \mathbb{F}_2^n
\) 
satisfying:  

\begin{itemize}
    \item For every $\alpha \in \mathbb{F}_2^n$ such that $|\widehat{f}(\alpha)| \geq \theta$, we have $\alpha \in \mathcal{S}$;
    \item For all $\alpha \in \mathcal{S}$, it holds that $|\widehat{f}(\alpha)| \geq \theta/2$.
\end{itemize}    

Having access to evaluations of $\chi_{\alpha_i}(x)$ for many inputs $x$, we can identify the characters in $\mathcal{S}$ up to a linear transformation. Let $\mathcal{T}$ be a subset of $\mathcal{S}$. Note that if 
\(
\sum_{\alpha_i \in \mathcal{T}} \alpha_i = \mathbf{0}^n,
\) 
then for all $x \in \mathbb{F}_2^n$,
\[
    \prod_{\alpha_i \in \mathcal{T}} \chi_{\alpha_i}(x) = \prod_{\alpha_i \in \mathcal{T}} (-1)^{\langle \alpha_i, x \rangle} = (-1)^{\left\langle \sum_{\alpha_i \in \mathcal{T}} \alpha_i , x \right\rangle} = 1.
\] 
Thus, the product of the corresponding columns of $Q$, denoted $\Pi_{\mathcal{T}}$, always equals $\mathbf{1}^\lambda$. On the other hand, if $\mathcal{B} \subseteq \mathcal{S}$ consists of linearly independent vectors, then the probability that 
$\prod_{\alpha_i \in \mathcal{B}} \chi_{\alpha_i}(x) = 1 \text{ for all } x$
is small.

\begin{claim}
\label{lemma_independence}
The probability that all entries of $\prod_{\alpha_i \in \mathcal{B}} \chi_{\alpha_i}(x_j)$ are equal to $1$ is at most $2^{-\lambda}$. Moreover, for any fixed $\mathcal{B}$, the probability that this product equals $1$ for any subset of $\mathcal{B}$ is at most $1/100$.
\end{claim}

\begin{proof}
Since the vectors $\alpha_i$ are linearly independent and the points $x_1, \dots, x_\lambda$ are sampled uniformly at random from $\mathbb{F}_2^n$, we have $\Pr[\prod_{\alpha_i \in \mathcal{B}} \chi_{\alpha_i}(x_j) = 1] = 1/2$ for each independent sample $x_j$. Hence, the probability that this holds across all $\lambda$ independent trials is at most $2^{-\lambda}$. Furthermore, the number of possible subsets of $\mathcal{B}$ is at most $2^{|\mathcal{B}|} = 2^{O(1/\theta^2)}$. By applying the union bound, the probability that $\prod_{\alpha_i \in \mathcal{D}} \chi_{\alpha_i}(x_j) = 1$ occurs for any subset $\mathcal{D} \subseteq \mathcal{B}$ is at most $2^{-\lambda} \cdot 2^{O(1/\theta^2)} < o(1)$, provided that $\lambda = \Omega((1/\theta^2)\log(1/\theta))$.
\end{proof}

Having identified all the heavy Fourier coefficients of $f$ (up to a linear transformation), we now estimate their corresponding Fourier magnitudes using the sample set $(Q, F)$. This is formalized in the following claim.

\begin{claim}
\label{lemm_approx_of_f2}
Using random examples $(Q, F)$, one can estimate all heavy Fourier coefficients of $f$ within $\pm \theta/10$, except with probability at most $1/25$.
\end{claim}

\begin{proof}
For each $B_i \in \{B_1, B_2, \dots, B_k\}$, the algorithm uses the random examples $(Q, F)$ to estimate $\widehat{f}(B_i)$ within an additive error of $\pm \theta/10$, achieving confidence $1 - \theta^2/100$. Let $\tilde{\widehat{f}}(B_i)$ denote this estimate. A sample size of $\lambda = (100/\theta^2) \log(100/\theta^2)$ suffices to achieve this accuracy and confidence. Applying the union bound over all heavy fourier coefficients, the probability that at least one estimate fails to meet the required accuracy is at most $1/25$.
\end{proof}

\begin{algorithm}
	\SetAlgoLined
 
	\textbf{Input:} Given a function $g$ such that $||\hat{g}||_1 \leq m$, and query access to $f$
    \\
        \textbf{Output:} $\dA(f, g) \leq \varepsilon$ or $\dA(f, g) > \varepsilon + \omega$ far from being linearly isomorphic
    \\
        \textbf{Initialization: }  $\theta = \frac{\omega}{10m}$, $\lambda = \frac{100}{\theta^2} \log \frac{100}{\theta^2}$ 
 
    \vspace{0.3cm}

 
        \textbf{Step 1:} $(Q, F) \gets \textbf{EconomicSieve}(\theta, \lambda)$
    
        \textbf{Step 2:} Relabel the columns of \( Q \) as \( \{B_1, \dots, B_k\} \), where \( B_1 = e_1^n, B_2 = e_2^n, \dots, B_r = e_r^n \), such that for all \( i \in \{r+1, \dots, k\} \), the column \( B_i \) is a linear combination of \( \{B_1, \dots, B_r\} \), where \( (e_i^n)_{i \in [r]} \) denote the standard basis vectors of \( \mathbb{F}_2^n \) along coordinate direction \( i \).
        
        \textbf{Step 3:} Use $(Q, F)$ to estimate each of $\widehat{f}(B_i)$ within $\pm \frac{\theta}{10}$ with confidence $1 - \frac{\theta^2}{100}$. Let $\tilde{\widehat{f}}(B_i)$ be the estimate

        \textbf{Step 4:} Define the function \( f^* \) such that \( \widehat{f^*}(\alpha) = \widetilde{\widehat{f}}(\alpha) \) for \( \alpha \in \{B_1, \dots, B_k\} \), and \( 0 \) otherwise, where \( \widetilde{\widehat{f}}(\alpha) \) denotes the estimated Fourier coefficient of \( f \) at point \( \alpha \in \mathbb{F}_2^n \).

        \textbf{Step 5:} Accept if and only if there exists a nonsingular linear transformation \( A \in \mathbb{F}_2^{n \times n} \) such that
        \[
            \sum_{\alpha \in \mathbb{F}_2^n} \widehat{g \circ A}(\alpha) \cdot \widehat{f^*}(\alpha) > 1 - 2\epsilon - \frac{\omega}{100}.
        \]


	\caption{\textbf{LinearIsoTester}}
	\label{algo: tester for lineariso}
\end{algorithm}

Finally, the algorithm constructs the real-valued function
\(
f^*(x) = \sum_{B_i} \widetilde{\widehat{f}}(B_i)\cdot \chi_{B_i}(x).
\)
The following claim shows that this approximation suffices for testing linear isomorphism, as the property is linear-invariant and $f^*$ preserves the spectral structure of $f$ up to a linear transformation.

\begin{claim}
\label{lem: distance of approximated coefficients}
Suppose $f$ and $g$ are $\epsilon$-close to being linearly isomorphic. Then there exists a nonsingular transformation $A$ such that
\(
\sum_{\beta} \widehat{f^*}(\beta)\cdot \widehat{g\circ A}(\beta) \geq 1 - 2\epsilon - \omega/100.
\)
Conversely, if there exists a nonsingular transformation $A$ satisfying
\(
\sum_{\beta} \widehat{f^*}(\beta)\cdot \widehat{g\circ A}(\beta) \geq 1 - 2\epsilon - \omega/100,
\)
then $f$ and $g$ are $\epsilon + \omega/100$-close to being linearly isomorphic.
\end{claim}

\begin{proof}
Since \( f \) and \( g \) are \( \epsilon \)-close to being linearly isomorphic, there exists an invertible linear transformation \( A \) such that
$
\Pr_{x \sim \mathbb{F}^{n}_{2}} \left[ f(x) \neq g \circ A(x) \right] \leq \epsilon.
$
Using the fact that \( f, g \in \{+1, -1\} \), we have:
\[
\sum_{\beta \in \mathbb{F}_2^n} \widehat{f}(\beta) \cdot \widehat{g \circ A}(\beta) = \mathop{\mathbb{E}}_{x \sim \mathbb{F}^{n}_{2}} \left[ f(x) \cdot g \circ A(x) \right] = 1 - 2 \cdot \mathop{\Pr}_{x \sim \mathbb{F}^{n}_{2}} \left[ f(x) \neq g \circ A(x) \right] \geq 1 - 2\epsilon.
\]
Since each Fourier coefficient of \( f^* \) approximates that of \( f \) within an additive error of at most \( \theta/10 \), we have:
\[
\sum_{\beta \in \mathbb{F}_2^n} \widehat{f^*}(\beta) \cdot \widehat{g \circ A}(\beta) \geq 1 - 2\epsilon - \frac{\theta}{10} \cdot \|\widehat{g}\|_1 \geq 1 - 2\epsilon - \frac{\omega}{100},
\]
where the last inequality follows from \( \theta = \frac{\omega }{10m} \) and \( \|\widehat{g}\|_1 \leq m \).

For the other direction, note that by assumption, the algorithm finds a transformation \( A \) such that
\[
\sum_{\beta \in \mathbb{F}_2^n} \widehat{f^*}(\beta) \cdot \widehat{g \circ A}(\beta) \geq 1 - 2\epsilon - \frac{\omega}{100}.
\]
Since each estimated Fourier coefficient in \( f^* \) is within \( \frac{\theta}{10} \) of the true value in \( \widehat{f} \), it follows that:
\[
\sum_{\beta \in \mathbb{F}_2^n} \widehat{f}(\beta) \cdot \widehat{g \circ A}(\beta) \geq 1 - 2\epsilon - \frac{\omega}{100} - \frac{\theta}{10} \cdot \|\widehat{g}\|_1 \geq 1 - 2\epsilon - \frac{2\omega}{100}.
\]

Using Parseval's identity, we have,
\begin{align*}
\sum_{\beta \in \mathbb{F}_2^n} \left( \widehat{f}(\beta) - \widehat{g \circ A}(\beta) \right)^2 
&= \sum_{\beta} \widehat{f}^2(\beta) - 2 \sum_{\beta} \widehat{f}(\beta) \cdot \widehat{g \circ A}(\beta) + \sum_{\beta} \widehat{g \circ A}^2(\beta) \\
&= 2 - 2 \sum_{\beta} \widehat{f}(\beta) \cdot \widehat{g \circ A}(\beta) \\
&\leq 2 - 2 \cdot \left(1 - 2\epsilon - \frac{2\omega}{100}\right) \\
&= 4\epsilon + \frac{4\omega}{100}.
\end{align*}
Further, 
\[
\Pr_{x \sim \mathbb{F}^{n}_{2}} \left[ f(x) \neq g \circ A(x) \right] = \frac{1}{4} \cdot \mathop{\mathbb{E}}_{x \sim \mathbb{F}^{n}_{2}} \left[ (f(x) - g \circ A(x))^2 \right] \leq \frac{1}{4} \cdot \left( 4\epsilon + \frac{4\omega}{100} \right) = \epsilon + \frac{\omega}{100}.
\]
Thus, \( f \) and \( g \) are \( \epsilon + \frac{\omega}{100} \)-close to being linearly isomorphic, as claimed.
\end{proof}

Finally, the overall failure probability of Algorithm~\ref{algo: tester for lineariso} is at most $1/10$ (Step 1) + $o(1)$ (Step 2) + $1/25$ (Step 3), which sums to at most $1/3$. Regarding the query complexity, note that the only queries to the unknown function $f$ occur in Step 1 via the \textsf{Economical Sieve}. By Lemma~\ref{lemma: economicsieve}, for the chosen parameters $\theta$ and $\lambda$, the total query complexity is $\tilde{O}(m^4/\omega^4)$. This completes the proof.
\end{proof}

\subsection{Proof of Lemma~\ref{lemma: economicsieve}}

The primary tool we use here is the projection of the Fourier coefficients of \( f \) onto a random coset structure and analyzing the concentration of these coefficients within the cosets. Below, we briefly outline the concept of random coset decomposition and then proceed to the main proof.

\vspace{1em}

Specifically, we project the Fourier spectrum of \( f \) onto a collection of cosets
\( \mathscr{C} \), obtained from a randomly permuted coset decomposition of a subspace
of codimension
\[
t \;=\; \left\lceil \log\!\left(\frac{2^{32}}{\theta^8}\right) \right\rceil .
\]
To construct this decomposition, we sample \( t \) independent uniform vectors
\( \beta_1, \ldots, \beta_t \in \mathbb{F}_2^n \) and define
\[
H \;=\; \mathrm{span}\{\beta_1, \ldots, \beta_t\}^{\perp},
\]
the subspace of vectors orthogonal to all \( \beta_i \).
Each coset of \( H \) is indexed by a vector \( b \in \mathbb{F}_2^t \) and defined as
\[
D(b)
\;:=\;
\bigl\{
\alpha \in \mathbb{F}_2^n
:\;
\langle \alpha, \beta_i \rangle = b_i
\text{ for all } i \in [t]
\bigr\}.
\]

To randomize the coset labels, we further sample a uniform shift
\( z \in \mathbb{F}_2^t \) and relabel each coset \( D(b) \) as \( D(b+z) \),
yielding a randomly permuted coset structure. As shown by Gopalan et al.~\cite{gopalan2011testing}, this construction induces a
pairwise independent hash family over \( \mathbb{F}_2^n \). We will rely heavily on this property throughout the proof. We now proceed to give the proof of Theorem \ref{lemma: economicsieve} by analyzing Algorithm~\ref{algo: economicsieve}. Throughout the proof, we will use the notations listed in Table~\ref{tab: notation}.

\begin{table}
    \centering
    \captionsetup{skip=6pt}
    \setlength{\arrayrulewidth}{0.25mm}
    \renewcommand{\arraystretch}{1.250}
    \addtolength{\arraycolsep}{6pt}
    \caption{Notations}
    \begin{tabular}{c | l}
        \hline
        \textbf{Notation} & \textbf{Description} \\
        \hline
        $\mathscr{C}$ & A random coset decomposition of $\mathbb{F}_2^n$ \\
        $C$ & Individual cosets of $\mathscr{C}$ \\
        $\mathcal{C}$ & Set of cosets of $\mathscr{C}$ that contains a {\em heavy} Fourier coefficient of $f$\\
        $\alpha_C$ & The Fourier coefficient in $C$ with the highest absolute value $\alpha_C \;:=\; \arg\max_{\beta \in C} \bigl|\widehat f(\beta)\bigr|$ \\
        $\mathcal{W}_C$ & Total Fourier weight of coset $C$ e.g. $\sum_{\beta \in C} \widehat f(\beta)^2$ \\
        $\mathcal{W}_C^*$ & $\widehat{f}^2(\alpha_C)$, weight of the heaviest Fourier coefficient in $C$ \\
        $\mathcal{P}_C(z)$ & $\sum_{\beta \in C} \widehat{f}(\beta) \chi_\beta (z) $, projection of $f(z)$ into coset $C$ \\
        $\mathcal{P}_C^*(z)$ & $\widehat{f}(\alpha_C )\chi_{\alpha_C}(z)$, projection of $f(z)$ onto the heaviest Fourier coefficient of $C$\\
        \hline
    \end{tabular}
    \label{tab: notation}
\end{table}

\begin{proof}[Proof of Lemma \ref{lemma: economicsieve} (\textsf{Economical Sieve})] 
In \textbf{Step 1}, we project the Fourier spectrum of \( f \) onto a random coset structure \( \mathscr{C} \) of codimension $\log \frac{2^{32}} {\theta^8} $. In \textbf{Step2}, we estimate the Fourier weight of \( f \) of each coset of \( \mathscr{C} \) within $\pm \frac{\theta^2}{4}$ accuracy. Given the parameter settings in the algorithm, we demonstrate that at the end of \textbf{Step 3} of Algorithm~\ref{algo: economicsieve}, we successfully identify a set of cosets \( \mathcal{C} \subseteq \mathscr{C} \) that collectively contain all heavy Fourier coefficients while ensuring that its size remains bounded. One may observe a resemblance of this guarantee to the celebrated Goldreich-Levin theorem \cite{goldreich1989hard}. This resemblance is not coincidental; in fact, the this part of this algorithm can be viewed as an implicit version of the Goldreich-Levin algorithm. We now formally state and prove the following claim.

\begin{claim}\label{lemma_bucket_discard}
Let $\mathscr{C}$ be a randomly permuted coset structure of codimension
$\log \frac{2^{32}}{\theta^8}$. Then, except with probability at most $\frac{1}{20}$, after
\textbf{Step~3} the algorithm outputs a set of cosets
$\mathcal C\subseteq\mathscr C$ such that:

\begin{enumerate}[label=(\roman*)]

    \item Every Fourier coefficient $\alpha$ with $|\widehat f(\alpha)|\ge \frac{\theta^2}{64}$ is mapped to a distinct coset. Consequently for any $C\in\mathscr C$  
    \[
        \max_{\beta \in C \setminus \{\alpha_c\}}
        \bigl|\widehat{f}(\beta)\bigr|
        \;<\;
        \frac{\theta^2}{64}
        \text{ where }
        \alpha_C \;:=\; \arg\max_{\beta \in C} \bigl|\widehat f(\beta)\bigr|
\]
    \item For any $C\in\mathscr C$, if $|\widehat f(\alpha_C)|\ge\theta$,
    then $C\in\mathcal C$.
    \item For every $C\in\mathcal C$, $|\widehat f(\alpha_C)|\ge \frac{\theta}{2}$. 
    
    \item For every $C\in\mathcal C$, its Fourier weight $\mathcal W_C = \sum_{\beta \in C} \widehat f(\beta)^2$ is highly concentrated around weight of its dominant fourier coefficient
 \[   
\sum_{\beta\in C\setminus\{\alpha_C\}} \widehat f(\beta)^2
<
\frac{\theta^4}{2^{12}}
\]
    
\end{enumerate}

\end{claim}

\begin{proof}

We begin by defining the following events.

\medskip
\noindent
\textbf{Event $E_1$:} All Fourier coefficients $\alpha$ with $|\widehat f(\alpha)|\ge \frac{\theta^2}{64}$ are mapped to distinct cosets of $\mathscr C$.

\medskip
\noindent
\textbf{Event $E_2$: } For every coset $C$ whose dominant Fourier coefficient $\alpha_C$ satisfies $|\widehat f(\alpha_C)| < \frac{\theta^2}{64}$,
\[
\mathcal W_C < \frac{\theta^2}{16}.
\]

\medskip
\noindent
\textbf{Event $E_3$:}
For every coset $C$ whose dominant coefficient satisfies $ |\widehat f(\alpha_C)| \ge \frac{\theta^2}{64}$,
\[
\mathcal W_C \le \widehat f(\alpha_C)^2 + \frac{\theta^4}{2^{12}}.
\]

\medskip
\noindent
\textbf{Event $E_4$: }
For every coset $C\in\mathscr C$, the estimate
$\widetilde{\mathcal W}_C$ produced in Step~3 satisfies
\[
|\widetilde{\mathcal W}_C - \mathcal W_C| \le \theta^2/4.
\]

\medskip
Condition on the event
$
E := E_1 \cap E_2 \cap E_3 \cap E_4.
$
we show that the conclusion of the claim follows deterministically.

\bigskip
\textbf{(i) Distinctness of heavy coefficients: } Since $\theta > \frac{\theta^2}{64}$, event $E_1$ immediately implies that all Fourier coefficients of magnitude at least $\theta$ are mapped to distinct cosets, establishing the first item of the claim.

\bigskip
\textbf{ (ii) All Cosets with heavy coefficients are selected: } Let $C$ be a coset whose dominant coefficient $\alpha_C$ satisfies $|\widehat f(\alpha_C)|\ge \theta$. Then
$
\mathcal W_C \ge \widehat f(\alpha_C)^2 \ge \theta^2.
$
By event $E_4$,
$
\widetilde{\mathcal W}_C
\ge
\mathcal W_C - \frac{\theta^2}{4}
\ge
\frac{3\theta^2}{4}.
$
Hence, $C$ is selected into $\mathcal C$.

\bigskip

\textbf{(iii) No coset with small dominant coefficient is selected: } Let $C$ be a coset whose dominant coefficient satisfies $|\widehat f(\alpha_C)|\le \theta/2$.

\medskip
\noindent
\emph{Case 1: $|\widehat f(\alpha_C)|\le \frac{\theta^2}{64}$.}

By event $E_2$,
$
\mathcal W_C < \frac{\theta^2}{16}.
$
Using event $E_4$,
$
\widetilde{\mathcal W}_C
\le
\mathcal W_C + \frac{\theta^2}{4}
<
\frac{3\theta^2}{4},
$
and hence $C$ is not selected.

\medskip
\noindent
\emph{Case 2: $\frac{\theta^2}{64} \le |\widehat f(\alpha_C)| \le \theta/2$.}

By event $E_3$,
$
\mathcal W_C
\le
\widehat f(\alpha_C)^2 + \frac{\theta^4}{2^{12}}
<
\frac{5\theta^2}{16}.
$
Again using event $E_4$,
$
\widetilde{\mathcal W}_C
\le
\mathcal W_C + \frac{\theta^2}{4}
<
\frac{3\theta^2}{4},
$
and $C$ is not selected.

\bigskip
\textbf{(iv) Fourier weight of the cosets is highly concentrated around the weight of the heaviest part:}
This follows directly from Event~\(E_3\).

\bigskip

It remains to show that
\[
\Pr_{H, b}[\neg E]
=
\Pr_{H, b}[\neg E_1 \cup \neg E_2 \cup \neg E_3 \cup \neg E_4]
<
\frac{1}{10},
\]
When the Fourier spectrum of \(f\) is projected onto a randomly permuted coset
structure by choosing a subspace \(H \le \mathbb{F}_2^n\) uniformly at random
of codimension
\( t = \log\!\left\lceil \frac{2^{12}\cdot 100}{\theta^8} \right\rceil\),
and an independent uniformly random shift \(b \in \mathbb{F}_2^t\), and
considering the restriction of \(f\) to the cosets \(b+H\). We analyze the events $E_1,E_2,E_3,E_4$ individually and bound the probability
that each of them fails.

\bigskip
\textbf{Event $E_1$ (Isolation of heavy coefficients).}
First we show that since the coset structure is induced by a uniformly random subspace of codimension \(t\),
any fixed vector lies in a given coset with probability \(2^{-t}\). To see why, fix \(\alpha \in \mathbb{F}_2^n\) and \(b \in \mathbb{F}_2^t\).
By definition, the event \(\alpha \in D(b+z)\) is equivalent to
\[
\forall i \in [t], \quad \langle \alpha, \beta_i \rangle = b_i + z_i .
\]
Each \(z_i\) is an independent uniformly random bit, and hence
\[
\Pr[\alpha \in D(b+z)] = \prod_{i=1}^t \Pr\!\left[z_i = \langle \alpha, \beta_i \rangle - b_i\right]
= 2^{-t}.
\]

Now consider two distinct vectors \(\alpha, \alpha' \in \mathbb{F}_2^n\).
They lie in the same coset if and only if
\[
\langle \alpha - \alpha', \beta_i \rangle = 0
\quad \forall i \in [t].
\]
Since \(\alpha - \alpha' \neq 0\), each constraint holds independently with
probability \(1/2\), and thus the collision probability is \(2^{-t}\). 

\medskip

Now, fix two distinct Fourier coefficients \(\alpha, \beta \in S\).
By the above, the probability that they collide into the same coset is \(2^{-t}\).
Applying the union bound over all \(\binom{|S|}{2}\) pairs, the total collision
probability is at most
\[
\binom{|S|}{2} 2^{-t}
\;\le\;
|S|^2 2^{-t}
\;\le\;
\left(\frac{2^{12}}{\theta^4}\right)^2 \cdot \frac{\theta^8}{2^{32}}
\;<\;
\frac{1}{100},
\]
where we used \(t = \left\lceil \log\!\left(2^{32}/\theta^8\right) \right\rceil\). Therefore, with probability at least \(0.99\), all coefficients in \(S\) are isolated
into distinct cosets.

\bigskip

\textbf{Event $E_2$ (Fourier Weight Concentration for Heavy cosets).}
Fix a coset $C$ containing a unique coefficient $\alpha_C\in S$.
Write
\[
\mathcal W_C
=
\widehat f(\alpha_C)^2
+
\sum_{\beta\neq\alpha_C}\widehat f(\beta)^2 I_\beta,
\]
where $I_\beta$ is the indicator that $\beta$ hashes to $C$. 

For each $\beta\neq\alpha_C$,
$
\mathop{\mathbb{E}}_{H,b}[I_\beta]=2^{-t}.
$
Therefore,
\[
\mathop{\mathbb{E}}_{H,b}\!\left[\mathcal W_C-\widehat f(\alpha_C)^2\right]
=
\sum_{\beta\neq\alpha_C}\widehat f(\beta)^2 \mathbb E_{H,b}[I_\beta]
=
2^{-t}\sum_{\beta\neq\alpha_C}\widehat f(\beta)^2.
\le 2^{-t}
\le \theta^8/2^{32}.
\]
Applying Markov’s inequality,
\[
\Pr_{H,b}\!\left[
\mathcal W_C-\widehat f(\alpha_C)^2 \ge \frac{\theta^4}{2^{12}}
\right]
\le
\frac{\theta^8/2^{32}}{\theta^4/2^{12}}
=
\theta^4/2^{20}.
\]
Since there are at most $|S|\le 2^{12}/\theta^4$ such cosets, a union bound gives
overall failure probability at most $1/100$.

\bigskip

\textbf{Event $E_3$ (Bounding Fourier Weights of Light cosets). }
Let $C$ be a coset such that $|\widehat f(\beta)|<\theta^2/64$ for all $\beta\in C$.
Then
\[
\mathop{\mathbb{E}}_{H,b}[\mathcal W_C]
=
\sum_{\beta}\widehat f(\beta)^2 \mathbb E_{H,b}[I_\beta]
=
2^{-t}\sum_{\beta}\widehat f(\beta)^2
\leq
2^{-t}
\le \theta^8/s^{32}.
\]
Using pairwise independence,
\[
\Var_{H,b}(\mathcal W_C)
=
\sum_{\beta}\widehat f(\beta)^4 \Var_{H,b}(I_\beta)
\le
\sum_{\beta}\widehat f(\beta)^4 \mathbb E_{H,b}[I_\beta]
=
2^{-t}\sum_{\beta}\widehat f(\beta)^4.
\]
Since $|\widehat f(\beta)|<\theta^2/64$ implies $\widehat f(\beta)^4\le \frac{\theta^4}{2^{12}} \widehat f(\beta)^2$,
\[
\sum_{\beta}\widehat f(\beta)^4
\le
\frac{\theta^4}{2^{12}}\sum_{\beta}\widehat f(\beta)^2
=
\frac{\theta^4}{2^{12}}
\]
Thus
$
\Var_{H,b}(\mathcal W_C)\le 2^{-t} \frac{\theta^4}{2^{12}}\le \frac{\theta^{12}}{2^{44}}.
$
Applying Chebyshev’s inequality,
\[
\Pr_{H,b} \left[ \mathcal W_C-\mathbb E_{H,b}[\mathcal W_C]\ge \frac{\theta^{2}}{16} \right]
\le
\frac{(\theta^{12}/2^{44})}{(\theta^{4}/16^{2})}=
\theta^8/2^{36}.
\]
There are at most $2^t\le 2^{32}/\theta^8$ cosets, so by a union bound the total
failure probability is at most $1/16$.

\bigskip

\textbf{Event $E_4$ (Fourier weight estimation are correct upto $\pm \theta^2/4$). }
We assume that it happens with probability at least $1-o(1)$. See Claim~\ref{lemma: economicsieve_query_complexity} further details. 

\medskip

Combining the failure probabilities, we get $ \Pr_{H, b}[\neg E_1 \cup \neg E_2 \cup \neg E_3 \cup \neg E_4] < \frac{1}{10}. $
\end{proof}

\begin{algorithm}
	\SetAlgoLined
 
	\textbf{Input:} Threshold: $\theta$, Number of samples : $\lambda$, and Distance: $\varepsilon$
    \\
        \textbf{Output:} Coset Samples  of $f \circ A$, with respect to subspace $\mathcal{S}(\theta)$
    \\
        \textbf{Parameters: }  $\eta = \frac{\theta}{8}$, $\gamma = \log 100\lambda$, 
    \\

    \vspace{0.3cm}

        \textbf{Step 1:} Let $\mathscr{C}$ be a randomly permuted coset structure of a randomly chosen subspace of codimension $\log \frac{2^{32}}{\theta^8}$\;
    
        \textbf{Step 2:} \ForEach{$C \in \mathscr{C}$}{
                    Estimate weight of $C$ within $\pm \theta^2/4$ accuracy\;
            }
    
        \textbf{Step 3:} Discard any coset with estimated weight $\leq \frac{3}{4}\theta^2$; Let $\mathcal{C}$ be the set of surviving cosets\;
         
        \textbf{Step 4:} \For{$i \in \{ 1, 2, 3, \cdots, \kappa\}$}{
                Sample $x_i$ uniformly at random from $\mathbb{F}_2^n$ and set $F[x_i] \gets f(x_i)$\;
                Sample $\{ y_1, y_2, \dots, y_\gamma \}$ each uniformly at random from $\mathbb{F}_2^n$\;
                    \ForEach{$y_j$}{
                        \ForEach{$C \in \mathcal{C}$}{
                            Estimate $P_C f(y_j)$ and $P_C f(x_i + y_j)$, each within $\pm \frac{1}{8} \theta$ accuracy\;
                        }
                    }
                    Set $Q[x_i][C] \gets \underset{j}{\text{median}}\{\text{sign}(\mathcal{P}_C[f(y_j)]) \cdot \text{sign}(\mathcal{P}_C[f(x_i + y_j)])\}$\;
                }

        \textbf{Step 5:} Return (Q, F)

	\caption{\textsf{Economical Sieve}}
	\label{algo: economicsieve}
\end{algorithm}

Having implicitly identified all the heavy cosets associated with heavy Fourier coefficients, the next step is to evaluate
$
\chi_{\alpha_c}(x)
$
at a uniformly sampled point \( x \in \mathbb{F}_2^n \), for each coset \( C \in \mathcal{C} \).
Importantly, while each survived coset \( C \) is known to contain a unique dominant Fourier character \( \alpha_c \), the identity of \( \alpha_c \) itself remains unknown. From a coding-theoretic viewpoint, this task can be interpreted as a local list-correction problem for the first-order Reed--Muller code. Given oracle access to a corrupted codeword (represented here by the Boolean function \( f \)), the goal is to recover the value of all nearby codewords at a queried position without explicitly decoding them. To this end, for any coset \( C \), define
\[
\mathcal{P}_C(z)
\;:=\;
\sum_{\beta \in C} \widehat{f}(\beta)\,\chi_\beta(z).
\]
The following claim shows that a noisy variant of the standard self-correction procedure succeeds simultaneously for all survived cosets.

\begin{claim}
\label{lem:sign_preserved}
Let \( \mathcal{C} \) be the set of survived cosets. Then for any fixed $x \in \mathbb{F}_2^n$,
\[
\Pr_{ y \sim \mathbb{F}_2^n}\Bigg[
\forall\, C \in \mathcal{C},\;
\sign\big(\mathcal{P}_C(x+y)\big)
\cdot
\sign\big(\mathcal{P}_C(y)\big)
=
\chi_{\alpha_c}(x)
\Bigg]
\;\ge\;
\frac{7}{8}.
\]
\end{claim}

\begin{proof}

Before proceeding the proof of Claim~\ref{lem:sign_preserved}, we show that  for every coset \( C \), for uniformly sampled $z$, $\mathcal{P}_C(z)= \sum_{\beta \in C} \widehat{f}(\beta) \chi_\beta (z) $, the projection of $f(z)$ onto a coset $C$ is highly concentrated around $\mathcal{P}_C^*(z)= \widehat{f}(\alpha_c) \chi_{\alpha_c}(z)$. More formally, we establish the following concentration result.

\begin{claim}
\label{lem:projection_concentration}
Suppose $f:\ftwo^n \to \{-1,+1\}$, and let $C\in\mathscr C$ be a coset with
dominant Fourier coefficient $\alpha_C$ such that
\[
\sum_{\beta\in C\setminus\{\alpha_C\}} \widehat f(\beta)^2
<
\frac{\theta^4}{2^{12}}
\quad\text{,}\quad
|\widehat f(\alpha_C)| > \frac{\theta}{2}.
\quad\text{and}\quad
\max_{\beta \in C \setminus \{\alpha_c\}}
\bigl|\widehat{f}(\beta)\bigr|
\;<\;
\frac{\theta^2}{64}.
\]
Then
    \begin{align*}
        \mathop{\mathrm{Pr}}_{z \sim \mathbb{F}^{n}_{2}} \left[ \left| \mathcal{P}_C(z) - \mathcal{P}_C^*(z) \right| \geq \frac{\theta^2}{64} + \tau \right] \leq \frac{\theta^4}{2^{12} \tau^2}.
    \end{align*}
\end{claim}

\begin{proof}
To establish this, we first recall that Fourier characters act as pairwise independent hash functions \( \chi_\alpha : \mathbb{F}_2^n \to \{-1, +1 \} \). Indeed, when \( z \) is uniformly sampled from \( \mathbb{F}_2^n \), each bit \( z_i \) is independent and uniformly distributed in \( \{0,1\} \). For any nonzero \( \alpha \in \mathbb{F}_2^n \), we have:
\[
\mathop{\mathbb{E}}_{z \sim \mathbb{F}^{n}_{2}}[\chi_\alpha(z)] = \mathop{\mathbb{E}}_{z \sim \mathbb{F}^{n}_{2}}[(-1)^{\langle \alpha, z \rangle}] = \frac{1}{2}(1) + \frac{1}{2}(-1) = 0.
\]
Moreover, for distinct \( \alpha_1, \alpha_2 \in \mathbb{F}_2^n \), we have:
\[
\mathop{\mathbb{E}}_{z \sim \mathbb{F}^{n}_{2}} \left[ \chi_{\alpha_1}(z) \cdot \chi_{\alpha_2}(z) \right] = \mathop{\mathbb{E}}_{z \sim \mathbb{F}^{n}_{2}} [(-1)^{\langle \alpha_1 + \alpha_2, z \rangle}] = 0 = \mathop{\mathbb{E}}_{z \sim \mathbb{F}^{n}_{2}}[\chi_{\alpha_1}(z)] \cdot \mathop{\mathbb{E}}_{z \sim \mathbb{F}^{n}_{2}}[\chi_{\alpha_2}(z)],
\]
establishing pairwise independence. Next, we show that the expectation of \( \bigl(\mathcal{P}_C(z) -  \mathcal{P}_C^*(z)\bigr) \) is small, analyzing two distinct cases separately depending on whether \( \mathbf{0} \in C' \) or not, where \( C' = \{ C - \alpha_c \} \).

\bigskip

\noindent
\textbf{Case I: When \( C \) does not contain \( \mathbf{0} \) or \( \mathbf{0} \) is the leader of the coset \( C \).}
\[ \mathop{\mathbb{E}}_{z \sim \mathbb{F}^{n}_{2}}\left[ \mathcal{P}_C(z) - \mathcal{P}_C^*(z) \right] 
= \mathop{\mathbb{E}}_{z \sim \mathbb{F}^{n}_{2}}\!\left[\sum_{\beta \in C'} \widehat{f}(\beta)\chi_\beta(z) \right]  
= \sum_{\beta \in C'} \widehat{f}(\beta)  \cdot \mathop{\mathbb{E}}_{z \sim \mathbb{F}^{n}_{2}}[\chi_\beta(z)]  
= \sum_{\beta \in C'} \widehat{f}(\beta) \cdot 0 = 0.\]

\bigskip

\noindent
\textbf{Case II: When \( C \) contains \( \mathbf{0} \) and \( \mathbf{0} \) is not the leader of the coset \( C \).} 
\[\mathop{\mathbb{E}}_{z \sim \mathbb{F}^{n}_{2}} \left[ \mathcal{P}_C(z) - \mathcal{P}_C^*(z) \right]  
= \mathop{\mathbb{E}}_{z \sim \mathbb{F}^{n}_{2}}\!\left[\sum_{\beta} \widehat{f}(\beta)\chi_\beta(z) \right] + \widehat{f}(0)  
= \sum_{\beta} \widehat{f}(\beta) \cdot \mathop{\mathbb{E}}_{z \sim \mathbb{F}^{n}_{2}}[\chi_\beta(z)] + \widehat{f}(0)  
\leq \widehat{f}(0) = \frac{\theta^2}{64}.
\]

Next, we show that the variance of \( \bigl(\mathcal{P}_C(z) - \mathcal{P}_C^*(z)\bigr) \) is also small:

\[ \Var_{z \sim \mathbb{F}^{n}_{2}}\!\left[\sum_{\beta} \widehat{f}(\beta)\chi_\beta(z) \right]  
= \sum_{\beta \in C'} \Var_{z \sim \mathbb{F}^{n}_{2}} [\widehat{f}(\beta)\chi_\beta(z) ]  
\leq \sum_{\beta \in C'} \mathop{\mathbb{E}}_{z \sim \mathbb{F}^{n}_{2}} \left[ \left( \widehat{f}(\beta)\chi_\beta(z) \right)^2 \right]  
= \sum_{\beta \in C'} \widehat{f}(\beta)^2  
= \frac{\theta^4}{2^{12}}
\]
Applying Chebyshev’s inequality,  
\[\Pr_{z \sim \mathbb{F}^{n}_{2}} \left[ \left| \mathcal{P}_C(z) - \mathcal{P}_C^*(z) \right| > \frac{\theta^2}{64} + \tau \right]  
\leq \frac{\Var_{z \sim \mathbb{F}^{n}_{2}}[\mathcal{P}_C(z) - \mathcal{P}_C^*(z)]} {\tau^2} 
\leq \frac{\theta^4}{2^{12} \tau^2} \] 
This completes the proof.
\end{proof}

For our algorithm, we set
$
\tau \;=\; \frac{\theta}{8}.
$
Substituting this value yields
\[
\Pr_{z \sim \mathbb{F}^{n}_{2}}\Big[
\big|\mathcal{P}_C(z) - \mathcal{P}_C^{*}(z)\big|
>
\frac{\theta}{4}
\Big]
\;\le\;
\frac{\theta^2}{64}.
\]

Next, observe that since \( f : \mathbb{F}_2^n \to \pmone \), the number of Fourier coefficients of magnitude at least \( \theta/2 \) is at most \( 4/\theta^2 \).
As each survived coset contains a unique dominant Fourier coefficient of magnitude at least \( \theta/2 \), it follows that
$
|\mathcal{C}| \;\le\; \frac{4}{\theta^2}
$. Applying a union bound over all survived cosets, we obtain
\[
\Pr_{z \sim \mathbb{F}^{n}_{2}}\Big[
\forall\, C \in \mathcal{C},\;
\big|\mathcal{P}_C(z) - \mathcal{P}_C^{*}(z)\big|
\ge
\frac{\theta}{4}
\Big]
\;\le\;
\frac{1}{16}.
\]

Combining this event with the fact that
$
\bigl|\widehat{f}(\alpha_C)\bigr| \;\ge\; \frac{\theta}{2}
$
for every surviving coset \(C\), and using that for any \(x \in \mathbb{F}_2^n\) the sum
\(x+y\) is uniformly distributed over \(\mathbb{F}_2^n\) when
\(y\) is chosen uniformly at random from \(\mathbb{F}_2^n\),
we conclude that

\[
\Pr_{z \sim \mathbb{F}^{n}_{2}}\Bigg[
\forall\, C \in \mathcal{C},\;
\sign\big(\mathcal{P}_C(x+y)\big)
\cdot
\sign\big(\mathcal{P}_C(y)\big)
=
\chi_{\alpha_c}(x)
\Bigg]
\;\ge\;
\frac{7}{8}.
\]
However, the algorithm does not have access to the exact value of $\mathcal{P}_C(z)$; it can only estimate it with reasonable accuracy. In our setting, if the estimated value $\widetilde{\mathcal{P}}_C(z)$ lies within $\pm \frac{\theta}{8}$ of the true value, the preceding argument still holds.  
\color{black}

Moreover, to construct the pair $(Q, F)$ under the given parameter settings of $\lambda$ and $\alpha$, it is necessary to evaluate $\chi_{\alpha_c}(x)$ for a sufficiently large number of inputs $x$, for all cosets $C \in \mathcal{C}$. This requires enhancing the reliability of the self-correction procedure. To achieve this, we apply the standard median trick for error reduction: for each fixed input $x$, we perform multiple independent trials using different random choices of $y$, and report the median of the obtained outcomes. By a direct application of the Chernoff bound, taking $O(\log(1/\theta))$ independent samples suffices to amplify the success probability to at least $1 - 1/\mathrm{poly}(1/\theta)$, which is sufficient to apply a union bound over all surviving cosets and all inputs $x$.
\end{proof}

\begin{claim}
\label{lemma: economicsieve_query_complexity}
In Algorithm~\ref{algo: economicsieve}, the total number of queries made by the algorithm is bounded by \( \tilde{O} \left( \max\left( \lambda/\theta^2, 1/\theta^4 \right) \right) \).
\end{claim}

\begin{proof}
The unknown function \( f \) is queried primarily in the following two steps of Algorithm~\ref{algo: economicsieve}:

\begin{description}
    \item[Step 2:] In this step, the algorithm estimates the weight of each of the \( O\left(\frac{1}{\eta^4}\right) \) cosets with accuracy \( \pm \frac{\theta^2}{4} \) and failure probability at most \( O(\eta^4) \). Setting \( \eta = \frac{\theta}{8} \) and then, the total number of queries required to estimate the weights of all cosets with the specified accuracy and confidence is
    $
    \tilde{O}\left( \frac{1}{\theta^4} \right).
    $. Note that for any \( x \in \mathbb{F}_2^n \), the weight of a coset \( r+H \) can be estimated via
    \[
    \mathcal{W}_{r + H} = \mathop{\mathbb{E}}_{x \sim \mathbb{F}_2^n, z \sim H^\perp} \left[ \chi_r(z) f(x) f(x + z) \right].
    \]
    Moreover, a single batch of samples can be reused simultaneously to estimate the weight of all cosets.

    \item[Step 4:] In this step, the algorithm samples \( \lambda \) many \( x \)'s, and for each, runs \( \tilde{O}(\log(1/\theta^2)) \) independent trials. In each trial, for each coset, the projections
    \[
    \mathcal{P}_{r+H}f(x) = \mathop{\mathbb{E}}_{y \sim H^\perp} \left[ \chi_r(y) f(x+y) \right]
    \]
    are estimated within \( \pm \frac{\theta}{8} \) accuracy with failure probability at most \( O(\theta^2) \). For any fixed \( x \), the same batch of samples can simultaneously estimate the projections for all cosets. Then, the total number of queries in this step is also bounded by
    $
    \tilde{O}\left( {\lambda}/{\theta^2} \right).
    $
\end{description}

So, the overall query complexity of Algorithm~\ref{algo: economicsieve} is $\tilde{O} \left( \max\left( {\lambda}/{\theta^2}, {1}/{\theta^4} \right) \right)$.
\end{proof}

This completes the analysis of \textsf{Economical Sieve} (proof of Lemma~\ref{lemma: economicsieve}). 
\end{proof}

\section{A lower bound for testing linear isomorphism}
\label{section: testing lower bound}

We begin by defining a class of Boolean functions known as the \emph{Maiorana--McFarland} functions. 
These functions have a long history of applications in theoretical computer science, particularly in proving lower bounds. 
Notable examples include circuit lower bounds~\cite{paul1977, blum1984} and studies on the structural properties of Boolean functions relevant to complexity theory~\cite{nisan1992, sanyal2019fourier}. 
Beyond complexity theory, they play a fundamental role in symmetric-key cryptography, especially in the design of stream ciphers, where they serve as building blocks for achieving good confusion (captured by the Fourier or Walsh spectrum) and diffusion (captured by the autocorrelation spectrum); see~\cite{SM00} for further details. 
We now formally define Maiorana--McFarland functions.

\begin{defi}
Given positive integers $n$ and $r$ with $r \leq n$, the family of Maiorana--McFarland functions (originally introduced in~\cite{mcfarland1973}), denoted $\text{MM}_{r,n}$, consists of $n$-variable Boolean functions $f: \mathbb{F}_2^n \to \{-1, +1\}$ of the form
\[
    g(x,y) = (-1)^{\langle x,\, \varphi(y) \rangle}, 
    \quad (x,y) \in \mathbb{F}_2^r \times \mathbb{F}_2^{n-r},
\]
where $\varphi : \mathbb{F}_2^{n-r} \to \mathbb{F}_2^r$ is an arbitrary mapping. 
\end{defi}

A key property of Maiorana--McFarland functions that we exploit is that, when composed with suitable linear transformations, their Fourier sparsity is governed by the rank of the underlying transformation. 
We state this property formally below.

\begin{claim}
\label{redfunc}
    Let $n = r + \log r$, and let $\varphi : \mathbb{F}_2^{n-r} \to \mathbb{F}_2^r$ be a mapping whose image has cardinality $r$, with the image set linearly independent in $\mathbb{F}_2^r$ and $L$ is a linear transformation in $\mathbb{F}_2^{r \times r}$. Let
    \[
        g_L(x,y) = (-1)^{\langle Lx,\,\varphi(y)\rangle},
    \]
    
    Then the Fourier sparsity of $g_L$ is at most $\operatorname{rank}(L)\times r$.
\end{claim}

\begin{proof}
For $(u,v) \in \mathbb{F}_2^r \times \mathbb{F}_2^{n-r}$, the Fourier coefficient of $g_L$ is
\[
    \widehat{g_L}(u,v) \;=\; \mathop{\mathbb{E}}_{x \sim \mathbb{F}_2^r,\, y \sim \mathbb{F}_2^{n-r}} 
    \left[ (-1)^{\langle Lx,\,\varphi(y)\rangle + \langle u, x\rangle + \langle v, y\rangle} \right].
\]
Reordering the expectation yields
\[
    \widehat{g_L}(u,v) \;=\; \mathop{\mathbb{E}}_{y \sim \mathbb{F}_2^{n-r}} 
    \left[ (-1)^{\langle v, y\rangle}\cdot 
    \mathbb{E}_{x \sim \mathbb{F}_2^r} (-1)^{\langle L^T\varphi(y) + u,\, x\rangle} \right].
\]
Now
\[
    \mathop{\mathbb{E}}_{x \sim \mathbb{F}_2^r} \left[ (-1)^{\langle L^T\varphi(y) + u,\, x\rangle} \right]
    = \begin{cases}
        1 & \text{if } u = L^T\varphi(y), \\
        0 & \text{otherwise}.
    \end{cases}
\]
Hence
\[
    \widehat{g_L}(u,v) \;=\; \frac{1}{2^{\,n-r}} \sum_{\substack{y \in \mathbb{F}_2^{n-r} \\ L^T \varphi(y) = u}} (-1)^{\langle v, y\rangle}.
\]

Therefore $(u,v)$ can be in the Fourier support only if $u\in\operatorname{Im}(L^T\circ\varphi)$. Since $\varphi$ has $r$ distinct outputs that are linearly independent in $\mathbb{F}_2^r$, the set 
$\{L^T\varphi(y) : y\in\mathbb{F}_2^{n-r}\}$ spans a subspace of dimension at most $\operatorname{rank}(L)$. Thus there are at most $\operatorname{rank}(L)$ distinct $u$ values that can occur. For each such $u$, there are $2^{\,n-r}=r$ possible choices of $v$. Hence the total number of possible nonzero Fourier coefficients is at most $\operatorname{rank}(L)\cdot r$, establishing the claimed sparsity bound.
\end{proof}

\subsection{Proof of Theorem \ref{theorem: lineariso testing lowerbound}}
In this section, we prove Theorem \ref{theorem: lineariso testing lowerbound}, establishing a lower bound that is quadratically stronger than the previously known result from \cite{gopalan2011testing}.

\lowerboundtheorem*

\begin{proof}

We establish the lower bound via a reduction from the Approximate Matrix Rank problem in randomized communication complexity (Theorem~\ref{theorem: actual rank}).  
Alice receives $A \in \ftwo^{r \times r}$, Bob receives $B \in \ftwo^{r \times r}$, and it is promised that $\mathrm{rank}(C)$ for $C = A+B$ is either $r$ or $r/4$.  
Their goal is to determine $\mathrm{rank}(C)$ with minimal communication and shared randomness.

Let $g = g_I$ denote the reference function defined in Claim~\ref{redfunc}, where $I$ is the $r \times r$ identity matrix.  
Alice constructs a function $f_A$ from $A$, Bob constructs $f_B$ from $B$, and together they define
\(
    f = f_C = f_{A+B}.
\)
By Claim~\ref{redfunc}, if $\mathrm{rank}(C) = r$, then $\mathrm{supp}(\widehat{f}) = r^2$ and if $\mathrm{rank}(C) = r/4$, then $\mathrm{supp}(\widehat{f}) \le r^2/4$. Moreover, if $\mathrm{rank}(C) = r$, then $f$ is linearly isomorphic to $g$. On the other hand, the following Claim shows that when $\mathrm{rank}(C) = r/4$, the function $f$ is far from every such isomorphism.

\begin{claim}
\label{lemma: farness}
If $\mathrm{rank}(C) = r$, then $f$ is $\tfrac{1}{4}$-far from any Boolean function with Fourier sparsity at most $r^2/4$.  
\end{claim}

\begin{proof}
 Let \( h: \mathbb{F}_2^n \to \pmone \) be a \( \frac{r^2}{4} \)-Fourier sparse function. Then, 
\begin{align}
\label{eq: dist sparse}
    \Pr_{x \sim \mathbb{F}^{n}_{2}} \left[h(x) \neq f(x)\right] 
    &=\Pr_{x \sim \mathbb{F}^{n}_{2}} \left[h(x) \neq f(x)\right] \\
    &= \frac{1}{2} + \frac{1}{2} \mathop{\mathbb{E}}_{x \sim \mathbb{F}^{n}_{2}} \left[h(x) f(x) \right] \notag \\
    &= \frac{1}{2} + \frac{1}{2} \sum_{\alpha \in \mathbb{F}_2^n} \hat{h}(\alpha) \hat{f}(\alpha) \notag \\
    &= \frac{1}{2} + \frac{1}{2} \sum_{\alpha \in \mathrm{supp}(h)} \hat{h}(\alpha) \hat{f}(\alpha).
\end{align}
Recall that for Boolean function $f : \ftwo^{n} \to \{\pm 1\}$, 
$\mathrm{supp}(\widehat{f}):=\left\{ \alpha \in \ftwo^{n} \, : \, \widehat{f}(\alpha) \neq 0\right\}$.
Now applying Cauchy-Schwarz inequality, we get 
\begin{align}
\label{equ: cauchy}
    \left|\sum_{\alpha \in \mathrm{supp}(h)}  \hat{h}(\alpha) \hat{f}(\alpha) \right|  &\leq \sqrt{\sum_{\alpha \in \mathrm{supp}(\widehat{h})}  \hat{h}^2(\alpha)  \cdot \sum_{\alpha \in \mathrm{supp}(\widehat{h})} \hat{f}^2(\alpha)} 
    = \sqrt{\sum_{\alpha \in \mathrm{supp}(\widehat{h})} \hat{f}^2(\alpha)}
\end{align}
Note that \( h \) is a \( \frac{r^2}{4} \)-Fourier sparse Boolean function, that is, $|\mathrm{supp}(\widehat{h})| \leq \frac{r^{2}}{4}$.
Observe that, by the construction of function $f$ (see~Corollary~\ref{redfunc}), the absolute values of any two non-zero Fourier coefficients are equal and the Fourier support $\mathrm{supp}(\widehat{f}) = r^{2}$. Using the fact that $\sum_{\alpha \in \ftwo^{n}} \widehat{f}(\alpha)^{2} = 1$ (Parseval's identity), we get 
\begin{align}\label{eqn:bound_Fourier_weight}
    \sqrt{\sum_{\alpha \in \mathrm{supp}(\widehat{h})} \hat{f}^2(\alpha)} \leq \frac{1}{2}
\end{align}
Finally, plugging the bound from Equation~\eqref{eqn:bound_Fourier_weight} into Equation~\eqref{eq: dist sparse}, we conclude that
\[
\mathop{\Pr}_{x \sim \mathbb{F}^{n}_{2}} \left[h(x) \neq f(x)\right] 
\geq \frac{1}{2} + \frac{1}{2} \cdot \left(-\frac{1}{2}\right) 
= \frac{1}{4}.
\]
\end{proof}

Thus, distinguishing between the two rank cases reduces to distinguishing whether $f$ is isomorphic to $g$ or $\tfrac{1}{4}$-far from every isomorphism of $g$.  
Suppose there exists a tester $\mathbb{T}$ for linear isomorphism with respect to a function of spectral norm $m$, with query complexity $q(m, \tfrac{1}{4})$.  
In our case, the reference function is $g$ with $m = O(r)$. To simulate a query $(x,y) \in \ftwo^n$, Alice computes $f_A(x,y)$, Bob computes $f_B(x,y)$, and they exchange one bit each.  
Since
\[
    f_C(x,y) = (-1)^{\langle (A+B)x, \varphi(y) \rangle} 
             = f_A(x,y) \cdot f_B(x,y),
\]
they can compute $f(x,y)$ with two bits of communication per query.  Hence, simulating $\mathbb{T}$ requires at most $2q(m,\tfrac{1}{4})$ bits. However, by Fact~\ref{theorem: actual rank}, solving the promised rank problem requires $\Omega(r^2)$ bits.  
Therefore, \( q(m,\tfrac{1}{4}) = \Omega(r^2)\). Since $m = O(r)$ for $g$, we conclude that testing linear isomorphism with respect to $g$ requires $\Omega(m^2)$ queries.
\end{proof}

\section{Conclusion}
\label{section:conclusion}

A Boolean function \( f: \mathbb{F}_2^n \to \{-1,1\} \) is said to be affinely isomorphic to another function \( g: \mathbb{F}_2^n \to \{-1,1\} \) if there exist an invertible matrix \( A \in \mathrm{GL}_n(\mathbb{F}_2) \) and a vector \( b \in \mathbb{F}_2^n \) such that for all \( x \in \mathbb{F}_2^n \),
\[
g(Ax + b) = f(x).
\]
By adapting the techniques from the proofs of Theorem~\ref{theorem: lineariso testing upperbound} and Theorem~\ref{theorem: lineariso testing lowerbound}, we can establish analogous results for testing affine isomorphism between a known function and an unknown function given oracle access to its truth table.

\bibliographystyle{plain}
\bibliography{ref}


\end{document}